\documentclass[twocolumn,floatfix,superscriptaddress,longbibliography,PRB]{revtex4}
\usepackage{graphicx,amsfonts,amssymb,amsmath, hyperref}
\usepackage[applemac]{inputenc}
\usepackage{tikz}
\usepackage{amsbsy}
\usepackage{bm}
\usetikzlibrary{arrows,decorations.pathmorphing,backgrounds,positioning,fit,petri,arrows.meta,intersections}
\usepackage{bbold}
\usepackage{bm}
\usepackage{color}
\usepackage{dcolumn}
\usepackage{feynmf} 
\usepackage{graphics}
\usepackage{graphicx}
\usepackage{subfigure}
\usepackage{slashed}
\usepackage{tensor}
\usepackage{placeins}
\usepackage{natbib}
\usepackage{relsize}
\usepackage {ifsym}
\usepackage{diagbox}
\usepackage{blindtext}
\usepackage{tikz}
\usepackage{bm}
\usetikzlibrary{shapes.misc}
\newcommand{\croix}[1][]
{%
\begin{tikzpicture}[baseline=(textbox.base),inner sep=2pt]
\node[cross out,draw,text width=\dimexpr#1] (textbox) {\strut};
\useasboundingbox (textbox);
\end{tikzpicture}%
}

\newif\ifhyper
\hypertrue
\ifhyper
\hypersetup{
  citecolor = {green},
  colorlinks = {true}, 
  urlcolor = {blue} 
}
\fi

\newlength{\ldag}
\def\eps{\epsilon}

\def\be{\begin{equation}}
\def\ee{\end{equation}}
\def\bea{\begin{eqnarray}}
\def\eea{\end{eqnarray}}
\def\bse{\begin{subequations}}
\def\ese{\end{subequations}}
\def\bc{\begin{center}}
\def\ec{\end{center}}

\def\nonum{\nonumber}

\def\D{{\rm d}}

\newcommand{\comment}[1]{}

\catcode`,\active

\catcode`\,12

\allowdisplaybreaks

\begin{document}

\title{Flat phase of polymerized membranes  at  two-loop order}

\author{O. Coquand} 
\email{coquand@lptmc.jussieu.fr}
\affiliation{Sorbonne Université, CNRS, Laboratoire de Physique Théorique de la Matière Condensée, 75005 Paris, France}
\affiliation{Institut für Materialphysik im Weltraum, Deutsches Zentrum für Luft- und Raumfahrt, Linder Höhe, 51147 Köln, Germany}

\author{D. Mouhanna} 
\email{mouhanna@lptmc.jussieu.fr}
\affiliation{Sorbonne Université, CNRS, Laboratoire de Physique Théorique de la Matière Condensée, 75005 Paris, France}

\author{S. Teber} 
\email{teber@lpthe.jussieu.fr}
\affiliation{Sorbonne Université, CNRS,  Laboratoire de Physique Théorique et  Hautes Energies, LPTHE,  75005  Paris, France}


\begin{abstract}

We investigate  two complementary  field-theoretical models   describing   the flat phase of  polymerized -- phantom -- membranes by means  of a two-loop, weak-coupling,  perturbative approach performed  near  the   upper   critical dimension $D_{uc}=4$,   extending the one-loop  computation of Aronovitz and Lubensky [Phys. Rev. Lett. {\bf 60}, 2634 (1988)]. We derive the renormalization group   equations within the modified minimal substraction scheme, then  analyze the corrections coming from two-loop   with a particular attention paid  to the anomalous dimension and the asymptotic  infrared  properties  of the  renormalization group  flow.   We finally  compare our  results to those provided by   nonperturbative  techniques  used to investigate these  two  models.  

\end{abstract} 
\maketitle

\section{Introduction}

Fluctuating surfaces  are ubiquitous in physics (see, e.g.,  Refs.\cite{proceedings89,bowick01}).  One meets them  within the context of  high-energy physics \cite{polyakov87,david89,wheater94,david04},  initially  through  high-temperature expansions   of  lattice gauge theories,  then  in the large-$N$ limit of gauge theories, in  two-dimensional quantum gravity,  in string theory  as world-sheet of string and, finally,  in brane theory.    They also  occur  as a fundamental  object of   biophysics  where  surfaces -- called in this context   {\sl membranes}   -- constitute the building blocks   of living  cells   such as   erythrocyte \cite{schmidt93,bowick01}.  Last but not least,  fluctuating surfaces -- or membranes  --  have  provided,  in condensed matter physics,   an extremely suitable  model to describe  both qualitatively and quantitatively   sheets  of graphene \cite{novoselov04,novoselov05}  or graphenelike materials (see,  e.g., \cite{katsnelson12} and references therein).

Two types   of membranes   should be   distinguished  regarding   their critical or, more generally,  long-distance properties: fluid membranes and polymerized membranes \cite{nelson04,nelson02}.    The specificity of fluid  membranes  is  that  the molecules are essentially free to diffuse inside the structure.  The consequence of this lack of  fixed connectivity is the  absence of  elastic properties.   As a result,  the free energy of the membrane  depends only on  its  shape --  its curvature -- and not on  a  specific  coordinate system.   Early  studies  \cite{degennes82,peliti85,helfrich85} have shown that strong height -- out-of-plane --   fluctuations  occur in such systems   in such a way   that the normal-normal correlation functions exponentially  decay with the distance over a typical persistence  length $\xi\sim e^{4\pi \kappa/T}$ in a way similar  to  what happens in   the two-dimensional $O(N)$  model. As a consequence,  there is no  long-range orientational order in fluid membranes --    that  are always {\it crumpled}  --  in agreement with the Mermin-Wagner theorem \cite{mermin66}.

Polymerized -- or tethered -- membranes  are more remarkable. Indeed due to the fact that molecules are tied together through a potential,  they display  a  fixed internal connectivity giving rise   to   elastic -- shearing and stretching --   contributions to the free energy.   It has been substantiated  that, in these conditions,  the   coupling between the out-of-plane    and  in-plane  fluctuations   leads  to a drastic  reduction  of the former \cite{nelson87}. This  makes   possible the existence  of  a phase transition   between  a  disordered {\sl crumpled}  phase  at high temperatures and an  ordered  {\sl  flat}  phase   with long-range order between the normals  at  low-temperatures  \cite{aronovitz88,david88,guitter89, paczuski88,aronovitz89,proceedings89},   analogous  to that occurring in ferro/antiferro magnets -- see, however,  below.   Although  the nature -- first or second order -- of this crumpled-to-flat  transition is still under debate \cite{kownacki02,koibuchi04,koibuchi05,kownacki09,koibuchi14,essafi14,satoshi16,cuerno16}  and the mere existence of a crumpled phase for realistic,  i.e., self-avoiding \cite{bowick01b},  membranes  seems to be compromised, there is no doubt about  the existence of a stable  flat phase.

Let  us  consider a   $D$-dimensional membrane embedded in the   $d$-dimensional Euclidean space.  The  location of a point on   the membrane is realized  by means  of  a  $D$-dimensional  vector  ${\bf x}$ whereas  a configuration of the membrane in the Euclidean space is described  through the embedding  ${\bf x}\to {\bf R} ({\bf x})$ with ${\bf R}\in  \mathbb{R}^d$.  One  assumes the existence of a low temperature, flat phase, defined by   ${\bf R}^0({\bf x)}=({\bf x},{\bf 0}_{d_c})$ where ${\bf 0}_{d_c}$ is the null vector  of  co-dimension $d_c=d-D$  and one decomposes  the field $\bf{ R}$ into $\bf{ R}(\bm{x)}=[{\bf x}+{\bf u}({\bf x}), {\bf h}({\bf x})]$  where ${\bf u}$ and ${\bf h}$ represent $D$ longitudinal  -- phonon -- and   $d-D$ transverse -- flexural  -- modes, respectively.  The  action of  a  flat phase configuration  ${\bf R}$ is given by \cite{nelson87,aronovitz88,david88,aronovitz89,guitter88,guitter89}
 \begin{equation}
\begin{array}{ll}
S[{\bf R}]=\displaystyle \int \text{d}^Dx \hspace{-0.3cm}  & \displaystyle \left\{  {\kappa \over 2}\big(\Delta  {\bf R}   \big)^2 
+ {\lambda\over 2}\,  u_{ii}^2 +  {\mu}\,  u_{ij}^2 \right\}
\label{action}
\end{array}
\end{equation}
where $u_{ij}$   is the  strain  tensor that parametrizes  the fluctuations around the flat phase configuration ${\bf R}^0({\bf x)}$:
$u_{ij}={1\over 2}(\partial_{i}{\bf R}.\partial_{j}{\bf R}-\partial_{i}{\bf R}^0.\partial_{j}{\bf R}^0)={1\over 2}(\partial_{i}  {\bf R}.\partial_{j}{\bf R}- \delta_{ij})$.
In Eq.~(\ref{action}),   $\kappa$ is the bending rigidity constant whereas   $\lambda$ and $\mu$ are the Lamé  coefficients;  stability considerations require that $\kappa$,  $\mu$,  and the bulk modulus $B=\lambda+2 \mu/D$  be all {\sl  positive}.

The most remarkable fact arising from the analysis of   (\ref{action}) is that, in the flat phase, the normal-normal correlation functions  display  long-range order  from  the upper critical ($uc$) dimension $D_{uc}=4$  down  to the  lower critical ($lc$) dimension $D_{lc}<2 \  $\cite{aronovitz88,aronovitz89}.  While  in apparent contradiction with the  Mermin-Wagner theorem \cite{mermin66},  this  result can be explained in the following way. At long distances, or low momenta,  typically given by  $\displaystyle q\ll \sqrt{\mu/ \kappa},\sqrt{\lambda/\kappa}$, the  term  $\big(\Delta  {\bf u}   \big)^2$ in (1) can be neglected with respect to   the terms  of the type   $(\partial_i u_j)^2$  entering in  the strain tensor $u_{ij}$, which are, thus, promoted to the rank of kinetic terms  of the field  $u$ \cite{gornyi15}.   It  follows immediately  from    power counting  considerations that the  nonlinear term   in the phonon field ${\bf u}$ appearing  in   $u_{ij}$, i.e. $\partial_i u_k\partial_j u_k$,   is  irrelevant; it can thus also be discarded. Under these assumptions the stain tensor $u_{ij}$ is given by: 
\begin{equation}
u_{ij}\simeq  {1\over 2} \left[\partial_i u_j+\partial_j  u_i+ \partial_i   {\bf h} . \partial_j   {\bf h}  \right]\ . 
\label{stress}
\end{equation}
It follows that  action  (1)  is  now quadratic in the phonon field  ${\bf u}$ and   one can   integrate over  it  exactly.   This   leads to an  {\sl effective} action   depending  only on the flexural  field ${\bf h}$.   In Fourier space, this effective action  reads \cite{ledoussal92,ledoussal18}
\begin{flalign}
&S_{\text{eff}}[{\bf h}\,] = \frac{\kappa}{2}\,\int_{\bf k}  \,  k^4\, |{\bf h}(\bf k)|^2 + 
\nonum \\
&+\frac{1}{4}\,\int_{{\bf k}_1,{\bf k}_2,{\bf k}_3,{\bf k}_4}  \hspace{-1.2cm}{\bf h}({\bf k}_1) \cdot{\bf h}({\bf k}_2)\,
R_{ab,cd}({\bf q})\,k_1^a\,k_2^b\,k_3^c\,k_4^d\,\,{\bf h}({\bf k}_3) \cdot {\bf h}({\bf k}_4)\, ,
\label{actioneff}
\end{flalign}
where $\int_{\bf k}=\int d^D{ k}/(2\pi)^D$ and ${\bf q} ={\bf k}_1 + {\bf k}_2 = -{\bf k}_3 -{\bf k}_4$.  The fourth  order, ${\bf q}$-transverse  tensor,  $R_{ab,cd}({\bf q})$   is given  by  \cite{ledoussal92,ledoussal18}
\begin{equation}
R_{ab,cd}({\bf q}) =  {\mu\,(D\lambda + 2\mu)\over \lambda + 2\mu}  N_{ab,cd}({\bf q}) + \mu\,M_{ab,cd}({\bf q})
\label{R}
\end{equation}
where    one has defined   the two mutually orthogonal tensors: 
\begin{equation}
\begin{array}{ll}
&\hspace{-0.4cm} N_{ab,cd}({\bf q}) =\displaystyle {1\over D-1}\,P_{ab}^T({\bf q}) \,P_{cd}^{T}({\bf q})
\\
\\
&\hspace{-0.4cm}M_{ab,cd}({\bf q}) =\displaystyle  {1\over 2}\,\big[P_{ac}^{T}({\bf q})\,P_{bd}^{T}({\bf q}) + P_{ad}^{T}({\bf q})\,P_{bc}^{T}({\bf q}) \big] - N_{ab,cd}({\bf q})
\nonumber
\end{array}
\end{equation}
where $P_{ab}^{T}({\bf q})=\delta_{ab}-q_aq_b/{\bf q}^2$ is the transverse projector.  Note that,  in $D=2$,  the tensor $M_{ab,cd}$ vanishes identically and  the  effective action (\ref{actioneff}) is parametrized  by only one coupling constant  which turns out to be proportional to   Young's  modulus \cite{nelson87,ledoussal92,ledoussal18}: $K_0=4\mu(\lambda+\mu)/(\lambda+2\mu)$. The key  point is that  the momentum-dependent interaction (\ref{R})  is nonlocal and gives rise to  a phonon-mediated interaction between  flexural modes  which  is of the long-range kind. More precisely,  this interaction  contains terms   such that   the product $R(\vert {\bf x}-{\bf y}\vert) \vert {\bf x}-{\bf y}\vert^2$ is not an integrable function in $D=2$  as required by the Mermin-Wagner theorem \cite{mermin66}  (see \cite{coquand19b} for a  detailed discussion).   This flat phase is characterized  by power-law behaviour for the phonon-phonon and flexural-flexural modes    correlation functions  \cite{aronovitz88,guitter88,aronovitz89,guitter89}: 
\begin{equation}
G_{uu}(q)\sim q^{-(2+\eta_u)} \hspace{0.3cm} {\hbox{and}} \hspace{0.3cm}  G_{hh}(q)\sim {q^{-(4-\eta)}}     
\label{correlation}
\end{equation}
where  $\eta$ and $\eta_u$ are nontrivial anomalous dimensions. In fact,   it  follows from Ward identities associated with   the remaining partial rotation invariance  of (\ref{action})  -- see below --  that $\eta$ and $\eta_u$ are not independent quantities and one has    $\eta_u=4-D-2\eta$ \cite{aronovitz88,guitter88,aronovitz89,guitter89}. 
Interestingly,  Eq.~(\ref{correlation}) provide  also an implicit equation for the lower critical  dimension  $D_{lc}$   defined as the dimension below which there is no more distinction between phonon and flexural  modes.  One gets from Eq.~(\ref{correlation}):  $D_{lc}-2+\eta(D_{lc})=0$ \cite{peliti85,aronovitz88,aronovitz89}. 
It results from this expression that the lower critical dimension $D_{lc}$, as well as   the associated anomalous dimension $\eta(D_{lc})$,  are no longer  given  by a power-counting analysis around a Gaussian fixed  point, as it occurs for  the $O(N)$ model, but by a  nontrivial computation of fluctuations. This implies, in particular, that there is no well-defined perturbative  expansion  of the flat phase theory  near the lower critical dimension $D_{lc}$   based on  the study of  a nonlinear $\sigma$   --  hard-constraints --   model  \cite{david88}.   

On the  other hand,  the soft-mode, Landau-Ginzburg-Wilson,  model   (\ref{action})  does not suffer from the same kind  of pathology;  a standard,  $\epsilon$-expansion about the upper critical dimension $D_{uc}$  is feasible and  has been performed  at leading order, a  long-time ago, in the seminal works  of  Aronovitz et al. \cite{aronovitz88,aronovitz89} and Guitter et al. \cite{guitter88,guitter89}  who have determined the renormalization group (RG) equations  and the properties of the flat phase  near $D=4$.   This perturbative approach  faces, however,  several drawbacks  that explain  why it has not been   pushed forward until  now:   (i)  It  involves an intricate momentum and tensorial  structure of  the propagators and  vertices  that  render   the diagrammatic  extremely rapidly  growing in complexity with the order of perturbation \cite{coquand20b}.  (ii)   The dimension of  physical   membranes, $D=2$,   is  ``far away"  from $D_{uc}$. Clearly, high orders  of  the perturbative series, followed by  suitable resummation techniques, are needed to get quantitatively trustable results.  The difficulty of carrying such a task is, however, increased by the  first drawback.    (iii)   The massless theory is manageable with  current modern techniques whereas,   with the $1/q^4$--form   of the flexural mode  propagator   $G_{hh}$  in (\ref{correlation}),  one apparently faces the problem of dealing with  infrared  divergences.    (iv)   The use of the dimensional regularization and, more precisely,  the modified minimal substraction ($\overline{\rm MS}$)  scheme,  which is by far the most convenient one, can  enter  in conflict with  the $D$-dependence of  physical  quantities or properties  -- see below.

In this context, several  nonperturbative methods -- with  respect  to the  parameter $\epsilon=4-D$ -- have been  employed   in order to tackle the physics directly in $D=2$.  Among them,   the $1/d_c$   expansion have been early performed  at leading order    \cite{david88,guitter88,aronovitz89,guitter89,gornyi15}  and, very  recently,  at  next-to-leading order  \cite{saykin20}. An  improvement  of the $1/d_c$ approximation that consists  in replacing, within this last  approach,  the bare propagator and vertices by their dressed and screened counterparts  leads to  the so-called   self-consistent  screening approximation  (SCSA)  that   has also been  used  at  leading   \cite{ledoussal92,zakharchenko10,roldan11,ledoussal18} and  next-to-leading order  \cite{gazit09}.    Finally,  a technique working  in all dimensions   $D$ and $d$, called    nonperturbative renormalization group (NPRG)  -- see below --   has been  employed to investigate  various kinds of membranes  at leading order of the so-called derivative expansion  \cite{kownacki09,essafi11,essafi14,coquand16a,coquand18,coquand20} and within an approach  taking into account  the full derivative  dependence  of the action  \cite{braghin10,hasselmann11}. Therefore, within the whole spectrum of approaches used  to investigate the properties of the flat phase of membranes,  it is only for the weak-coupling perturbative approach  that the next-to-leading order is still missing (see however \cite{mauri20}).  This is clearly  a flaw as the subleading corrections of any approach generally provide  valuable insights  on  the structure of the whole theory.  They also convey  useful information  about the accuracy of  complementary approaches. 

 We  propose  here to  fill this gap  and to  investigate the properties of the  flat phase of  polymerized membranes at two-loop order in the coupling constants, near  $D_{uc}=4$,  considering successively the  flexural-phonon, {\sl two-field},  model  (\ref{action})  and then  the flexural-flexural,   {\sl effective} model   (\ref{actioneff}). We compute the  RG functions of these two models, analyze their fixed points and compute the corresponding anomalous dimensions. Finally,   we compare  these   results    together and, then,  with  those   obtained from nonperturbative methods.  Note that, due  to the length of  the computations and expressions involved,  we  restrict  here ourselves to the main  results; details will be given in a forthcoming publication \cite{coquand20b}.

\section{The two-field  model}

 \subsection{\bf  The perturbative approach}

We first consider  the two-field model (\ref{action})   truncated by means of  the long distance  approximations Eq.~(\ref{stress}) and $\big(\Delta  {\bf u}   \big)^2\simeq 0$.  The perturbative approach  proceeds as usual: one expresses the  action in terms of  the phonon and flexural  fields  ${\bf u}$ and ${\bf h}$  then get the propagators and  3 and 4-point  vertices, see \cite{guitter89, coquand20b}.   A crucial issue is that, although the truncations  of    action  (\ref{action})   above    break  its  original  $O(d)$ symmetry,   a partial rotation invariance remains \cite{guitter88,guitter89}: 
\begin{equation}
\begin{array}{l}
 \displaystyle   {\bf h}  \mapsto   {\bf h}  +{\bf A} _{ i}\, x_i\\   [0cm]
 u_i  \mapsto   u_i - {\bf A} _{i}. {\bf h}  - \displaystyle \frac{1}{2} {\bf A} _{i}.{\bf A} _{j}\, x_j
 \end{array}
 \nonumber
  \end{equation}
where ${\bf A}_i$ is any  set of  $D$ vectors $\in\mathbb{R}^{d_c}$.   From this property  follow  Ward identities for the effective action $\Gamma$ \cite{guitter88,guitter89}:
 \begin{equation}
 \int  d^D  x\bigg({\bf h}\frac{\delta \Gamma}{\delta u_i}-x_i\frac{\delta\Gamma}{\delta {\bf h}}\bigg)=0\  . 
 \label{ward}
  \end{equation}
One easily shows that this equation is solved by -- the truncated form of --   (\ref{action}),  thereby ensuring  the renormalizability of the theory.  Moreover, from  (\ref{ward}),  one can derive  successive identities  relating various   $n$-points  to $(n-1)$-point  functions in such a way that only  the renormalizations   of phonon and flexural modes  propagators  are    required. This is a tremendous simplification of the computation which, nevertheless, 
preserves   a nontrivial algebra.   Also, as previously mentioned, an  apparent  difficulty comes from the structure of the --  bare --  flexural mode  propagator  $G_{hh}(q)\sim 1/q^4$ and the  masslessness of the theory  that suggests  that  the perturbative expansion  could be plagued by severe infrared divergencies. In this respect,  one has first  to note  that  the masslessness of the theory and the form of the propagators  (\ref{correlation}) are somewhat  contrived as  they originate from the derivative character of (\ref{action})  
relying itself from the lack of  translational invariance of  the embedding ${\bf x}\to {\bf R} ({\bf x})$.  It appears that the {\sl  natural} objects   that should be ideally considered are  the tangent-tangent correlation functions $\widetilde G\sim \langle \partial_i {\bf R}.\partial_i {\bf R} \rangle$ whose Fourier transforms  are, for fixed-connectivity membranes,  proportional to the position-position ones  $G\sim \langle  {\bf R}. {\bf R} \rangle$  with  a factor of ${\bf q}^2$ \cite{aronovitz89} and  are, consequently,   infrared safe. In practice, however,   employing  the  latter correlation functions   is  both preferable and   innocuous as its use only implies the appearance of  tadpoles  that cancel order by order in perturbation theory.  One can, thus, proceed  using  dimensional regularization in the conventional way  ignoring the occurrence of  possible infrared poles  \cite{smirnov83}.  

 \subsection{\bf  The renormalization group equations} 

One   introduces the renormalized fields ${\bf h}_R$ and   ${\bf u}_R$ through $\displaystyle {\bf h}=Z^{1/2}\kappa^{-1/2}{\bf h}_R $ and   ${\bf u}=Z\kappa^{-1} {\bf u}_R$ and the renormalized coupling constants  $\lambda_R$ and   $\mu_R$ through :
\begin{equation}
\begin{split}
&\displaystyle \lambda=k^\epsilon Z^{-2}\kappa^2Z_\lambda \lambda_R
\\
&\mu=k^\epsilon Z^{-2}\kappa^2Z_\mu \mu_R 
\label{defren}
\end{split}
\end{equation}
where $k$ is the renormalization momentum scale  and $\epsilon=4-D$. Within the  $\overline{\rm MS}$  scheme,  one introduces the scale  $\overline{k}^2=4\pi e^{-\gamma_E}k^2$ where $\gamma_E$ is the Euler constant.   One then   defines the $\beta$-functions  $\beta_{\lambda_R}=\partial_t\lambda_R$ and $\beta_{\mu_R}=\partial_t\mu_R$,  with  $t=\ln \overline{k}$.  As usual, in  order to write  these quantities  in terms of the field and coupling constant renormalizations  $Z=Z(\lambda_R,\mu_R,\epsilon)$,   $Z_\lambda=Z_\lambda(\lambda_R,\mu_R,\epsilon)$ and $Z_\mu=Z_\mu(\lambda_R,\mu_R,\epsilon)$ one expresses the  independence of the bare coupling constants   $\lambda$ and $\mu$ with respect to $t$:  ${d\lambda/  dt} ={d\mu/  dt} =0$. Using   (\ref{defren})  and defining  the anomalous dimension 
\begin{equation}
\eta=\beta_{\lambda_R}\, {\partial\ln Z\over \partial \lambda_R}+\beta_{\mu_R}\, {\partial\ln Z\over \partial \mu_R}
\nonumber
\end{equation} 
one gets from these conditions:
\begin{equation}
\begin{split}
&\beta_{\lambda_R}\, \partial_{\lambda_R} \ln(\lambda_R Z_\lambda) +\beta_{\mu_R}\, \partial_{\mu_R}\ln(\lambda_R Z_\lambda)=-\epsilon+2\eta \\
\\
&\beta_{\lambda_R}\, \partial_{\lambda_R} \ln(\mu_R Z_\mu) + \beta_{\mu_R}\, \partial_{\mu_R}\ln(\mu_R Z_\mu)=-\epsilon+2\eta\ 
\nonumber
\end{split}
\end{equation}
where only simple poles in $\epsilon$ of  $Z$, $Z_\lambda$ and $Z_\mu$   have to be considered,  see \cite{coquand20b}.

Computations have been performed independently by means of  (i)  conventional renormalization --  counterterms --   method     (ii)    BPHZ \cite{Bogoliubov:1957gp,Hepp:1966eg,Zimmermann:1969jj} renormalization scheme  with the  help  of  the  LITERED  {\it mathematica} package for the reduction  two-loop integrals \cite{Lee:2013mka}.   Both computations  have required   techniques for computing  massless Feynman diagram calculations that are reviewed  in, e.g., Ref.\cite{Kotikov:2018wxe}.

Omitting the $R$ indices on the renormalized coupling constants  one gets, after  involved computations \cite{coquand20b}  
\begin{equation}
\begin{split}
\beta_{\mu}&=-\epsilon\mu + 2 \mu\, \eta+\frac{d_c\,\mu^2}{6(16\pi^2)}\bigg(1+\frac{227}{180}\, \eta^{(0)}\bigg) \\
\\
\beta_{\lambda}&=-\epsilon\lambda + 2\lambda\, \eta  +\frac{d_c\big(6\lambda^2+6\lambda\mu+\mu^2\big)}{6(16\pi^2)}\\
\\
&-\frac{d_c \big(378\lambda^2-162\lambda\mu-17\mu^2\big)}{1080\, (16\pi^2)}\, \eta^{(0)}
-\frac{d_c^2\,\mu(3\lambda+\mu)^2}{36(16 \pi^2)^2}
\end{split}
\label{2loops}
\end{equation}
where 
\begin{equation}
\begin{split}
&\eta=\eta^{(0)}+\eta^{(1)}=\frac{5\mu(\lambda+\mu)}{16\, \pi^2(\lambda+2\mu)}\\
\\
&-\frac{\mu^2\Big((340+39\,d_c)\lambda^2+4(35+39\,d_c)\lambda\mu+(81\,d_c-20)\mu^2\Big)}{72\,  (16 \pi^2)^2(\lambda+2\mu)^2}\ . 
\label{eta}
\end{split}
\end{equation}


 \subsection{\bf  Fixed points analysis}

Equations (\ref{2loops}) and (\ref{eta}) constitute  the first set of  our main results. These equations  extend to two-loop order  those of Aronovitz and Lubensky \cite{aronovitz88}.    One first  recalls  the properties of the one-loop RG  flow \cite{aronovitz88,guitter88,aronovitz89,guitter89},  then,   considers  the full  two-loop equations (\ref{2loops}) and (\ref{eta}).

\subsubsection{\bf One-loop order} 

\begin{figure}
\includegraphics[scale=0.28]{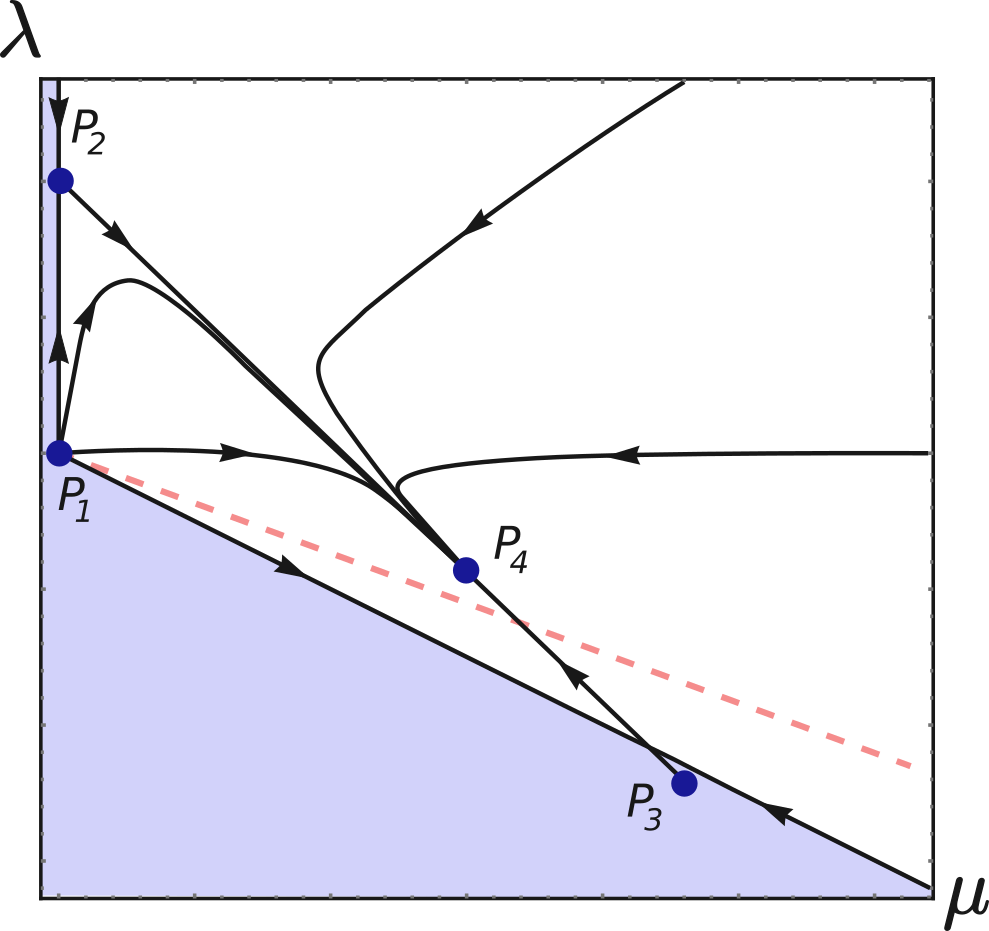}
\caption{The schematic RG  flow diagram (not to scale)  on the plane $(\mu,\lambda)$. The stability region of action (\ref{action}) is delimited by the line $2\lambda+\mu=0$ on which lies the fixed point $P_3$ at one-loop order and the line $\mu=0$ on which lies the fixed point $P_2$. The dashed line  corresponds to the one-loop attractive subspace   $3\lambda+\mu=0$ where   the  stable fixed point $P_4$ stands.  At two-loop order,  $P_4$  does not stand exactly on the line $3\lambda+\mu=0$ anymore whereas   $P_3$ is ejected out the stability region.}
\label{flow}
\end{figure}

At  one-loop order there are four  fixed points, see Fig.~\ref{flow}: 

\vspace{0.2cm}
(i)  the Gaussian one  $P_1$   for which $\mu^*_1=0,\, \lambda^*_1=0$ and $\eta_1=0$; it  is twice unstable. 

\vspace{0.2cm}

(ii)  The  -- shearless  --  fixed point  $P_2$  with $\mu^*_2=0,\, \lambda^*_2={16\pi^2\,\epsilon/d_c}$ and $\eta_2=0$ which lies on the  stability line $\mu=0$; it  is once  unstable.

\vspace{0.2cm}

(iii)   The infinitely compressible  fixed point $P_3$ with $\mu^*_3=96\pi^2\,\epsilon/(20+d_c), \, \lambda^*_3=-48\pi^2\,\epsilon/(20+d_c)$ and $\eta_3=10\,\epsilon/(20+d_c)$, for which   the bulk modulus $B$  vanishes,  i.e.\  $2\lambda^*_3+\mu^*_3=0$. It is  thus  located on the corresponding  stability line;  it is  once  unstable.  

\vspace{0.2cm}

(iv)  The flat phase fixed point   $P_4$  for which  $\mu^*_4=96\pi^2\,\epsilon/(24+d_c), \lambda^*_4=-32\pi^2\,\epsilon/(24+d_c)$ and  $\eta_4=12\,\epsilon/(24+d_c)$. It is   fully stable and, thus, controls the  flat phase at long distance. At one-loop order,  this  fixed point  is  located on the stable line $3\lambda+\mu=0$ -- that, in $D$ dimensions,  generalizes  to  the line $(D+2)\lambda+2\mu=0$. 


\subsubsection{\bf Two-loop order}

\vspace{0.2cm}

At two-loop order there are still four fixed points.   For the two first ones,   nothing changes whereas, for the two last ones,  the situation changes only marginally:  

\vspace{0.2cm}

(i)  the Gaussian fixed point   $P_1$  remains twice unstable. 

\vspace{0.2cm}

(ii)  The  once unstable fixed point  $P_2$  keeps  the same coordinates as at  one-loop order -- with in particular $\mu_2^*=0$ --     thus  the associated  anomalous dimension,  which is proportional to $\mu$, see (\ref{eta}),   still vanishes: $\eta_2=O(\eps^3)$. 

\vspace{0.1cm}

(iii)  At  the other once  unstable  fixed point $P_3$, whose  coordinates and associated exponent  are given in Table \ref{tablexpo3},   the bulk modulus $B$ becomes  now slightly negative -- and of order $\epsilon^2$ --  see Fig.\ref{flow}.  It follows that,  at this order,  $P_3$ is ejected out of the stability region.  {\sl However},  we emphasize that  this fact  fully depends on the  technique   or   -- two-field  or effective -- formulation of the  theory  -- see below. It is, thus,  likely that this is an artifact of the present computation.  So one can  still  consider  $P_3$   as potentially present in the genuine  flow diagram of membranes. 
\begin{table}[h]
\begin{center}
\begin{tabular}{|c|l|c|c|c|c|}
\hline
  \rule[-0.4cm]{0cm}{1.cm} $\mu^*_3$  &  $\displaystyle\frac{96\pi^2\, \epsilon}{20+d_c} +\frac{80\pi^2(-d_c+232)}{3(20+d_c)^3}\, \epsilon^2$            \\ 
\hline
  \rule[-0.4cm]{0cm}{1.cm}   $ \lambda^*_3$    & $\displaystyle\ -\frac{48\pi^2 \,\epsilon}{20+d_c}-\frac{8\pi^2(9 d_c^2+265\, d_c+2960)}{3(20+d_c)^3}\, \epsilon^2$             \\
 \hline
  \rule[-0.4cm]{0cm}{1.cm}  $\eta_3$        & $\displaystyle\ \frac{10\,\epsilon}{20+d_c}-\frac{d_c(37\,d_c+950)}{6(20+d_c)^3}\,\epsilon^2$    \\
  \hline
\end{tabular}
\end{center}
\caption{ Coordinates $\mu^*_3$ and $\lambda^*_3$ of the  fixed point  $P_3$  and the corresponding  anomalous dimension $\eta_3$  at order $\epsilon^2$ obtained from the two-field model.}
\label{tablexpo3}
\end{table}

(iv)  $P_4$ remains  fully stable and, thus,  still controls  the flat phase. Its  coordinates and associated anomalous dimension   are given in Table \ref{tablexpo4}.  As a noticeable point  one indicates that    this  fixed point    no longer lies on the  line $(D+2)\lambda+2\mu=(6-\epsilon)\lambda+2\mu=0$  -- with a distance of order $\epsilon^2$ as expected --  which is, thus, no longer  an attractive  line in the infrared. 

\vspace{0.1cm}

As can be seen  in Table \ref{tablexpo4}  the anomalous dimension at  $P_4$  is only  very slightly modified with respect to its  one-loop order value.  The extrapolation of our result for $\eta_4$   to $D=2$,   i.e.\  $\epsilon=2$ and $d_c=1$ leads, at one and two-loop orders,  to $\eta_4^{1l}=24/25=0.96$ and $\eta_4^{2l}=2856/3125\simeq 0.914$. These  values are obviously only indicative   and are in no way  supposed to provide  a quantitatively  accurate  prediction in $D=2$. However, one can   note    that the two-loop correction moves  the  value of $\eta_4$  towards  the   right direction if one refers to the  generally accepted numerical data that lie  in the range of $[0.72,0.88]$  \cite{guitter90,zhang93,bowick96,gompper97,los09,troster13,wei14,troster15,los16,kosmrlj17,hasik18}. 

\begin{table}[h]
\begin{center}
\begin{tabular}{|c|l|c|c|c|c|}
\hline
   \rule[-0.4cm]{0cm}{1.cm} $\mu^*_4$ &   $\displaystyle\frac{96\pi^2 \, \epsilon}{24+d_c}-\frac{32\pi^2(47 d_c+228)}{5(24+d_c)^3}\, \epsilon^2$          \\ 
\hline
  \rule[-0.4cm]{0cm}{1.cm}   $ \lambda^*_4$    &  $\displaystyle\ -\frac{32\pi^2 \,\epsilon}{24+d_c}+\frac{32\pi^2(19 d_c+156)}{5(24+d_c)^3}\, \epsilon^2$         \\
 \hline
  \rule[-0.4cm]{0cm}{1.cm}  $\eta_4$         & $\displaystyle\ \frac{12\,\epsilon}{24+d_c}-\frac{6 d_c(d_c+29)}{(24+d_c)^3}\,\epsilon^2$         \\
  \hline
\end{tabular}
\end{center}
\caption{ Coordinates $\mu^*_4$ and $\lambda^*_4$ of the flat  phase  fixed point  $P_4$  and the corresponding  anomalous dimension $\eta_4$ at order $\epsilon^2$ obtained from the two-field model.}
\label{tablexpo4}
\end{table}


\section{The flexural mode  effective model}

 \subsection{\bf  The perturbative approach}

 We have also  considered an alternative approach to the flat phase theory of membranes  which is given by  the flexural mode  effective model  (\ref{actioneff}). There  are three  main reasons   to tackle directly this model. The first one is formal and consists  in  showing  that one can treat, at two-loop order,  a model with a nonlocal interaction.  The second reason   is that  this    provides  a nontrivial check of  the previous computations.   Indeed the field-content,  the (unique) four-point nonlocal  vertex  as well as the whole structure of the perturbative expansion  of the effective model (\ref{actioneff}) are considerably different from those  of the two-field model so that the agreement between the two approaches is a  very substantial fact. The last  reason to investigate this model  is that it  involves  a new  coupling constant   $b= {\mu\,(D\lambda + 2\mu)/(\lambda + 2\mu)}$, see (\ref{R}),   which: (i)   is directly  proportional to the bulk modulus $B$  associated  with a stability line of the model and (ii) incorporates   a $D$-dependence which, as $b$ is considered  as a  coupling constant in itself, will  be kept from the  influence of the dimensional regularization. 

 \subsection{\bf  The renormalization group equations} 

As in the two-field model,  one introduces the renormalized field  ${\bf h}_R$  through   $\displaystyle {\bf h}=Z^{1/2}\kappa^{-1/2}{\bf h}_R $, the renormalized coupling constants  $b_R$ and   $\mu_R$ through 
\begin{equation}
\begin{split}
&\displaystyle b=k^\epsilon Z^{-2}\kappa^2Z_b\; b_R
\\
&\mu=k^\epsilon Z^{-2}\kappa^2Z_\mu\; \mu_R
\label{defren2}
\end{split}
\end{equation}
and the $\beta$-functions  $\beta_{b_R}=\partial_t b_R$ and $\beta_{\mu_R}=\partial_t\mu_R$.  Using (\ref{defren2}) 
to express the independence of the bare coupling constants $b$ and $\mu$ with respect to $t$ and  defining  the   anomalous dimension 
\begin{equation}
\eta=\beta_{b_R}\, {\partial\ln Z\over \partial b_R}+\beta_{\mu_R}\, {\partial\ln Z\over \partial \mu_R}
\nonumber
\end{equation}
the $\beta$-functions  $\beta_{b_R}$ and $\beta_{\mu_R}$  read:
\begin{equation}
\begin{split}
&\beta_{b_R}\, \partial_{b_R} \ln(b_R Z_b) +\beta_{\mu_R}\, \partial_{\mu_R}\ln(b_R Z_b)=-\epsilon+2\eta \\
\\
&\beta_{b_R}\, \partial_{b_R} \ln(\mu_R Z_\mu) +\beta_{\mu_R}\, \partial_{\mu_R} \ln(\mu_R Z_\mu)=-\epsilon+2\eta\ . 
\nonumber
\end{split}
\end{equation}

After a rather heavy algebra and using the same techniques as for  the two-field  model  one  gets:
\begin{equation}
\begin{split}
&\beta_{\mu}=-\epsilon\mu + 2 \mu\,  \eta+\frac{d_c\,\mu^2}{6 (16\pi^2)}\bigg(1+\frac{107  b +574\,\mu}{216\,  (16 \pi^2)}\bigg) \\
\\
&\beta_b=-\epsilon b + 2 b\,  \eta\ +\frac{5 d_c\, b^2}{12(16\pi^2)}\bigg(1+\frac{178\, \mu-91 b}{216\,  (16 \pi^2)}\bigg) 
\end{split}
\label{2loopseff}
\end{equation}
and: 
\begin{equation}
\begin{split}
\eta&=\frac{5(b+2\mu)}{6(16\pi^2)}  + \\
& \frac{5\,(15\,d_c - 212)\,b^2 + 1160\,b\,\mu - 4\,(111\,d_c - 20)\,\mu^2}{2592 (16\pi^2)^2}\, .
\label{etaeff}
\end{split}
\end{equation}

\subsection{\bf  Fixed point  analysis} 

Equations (\ref{2loopseff}) and (\ref{etaeff}) constitute our  second   set  of  results.  We now analyze their content.  

\subsubsection{\bf One-loop order}

 At  one-loop  one finds  four  fixed points: 
 
 \vspace{0.2cm}
 
 (i)  the Gaussian one  $P_1$ with $ \mu^*_1=0,\, b^*_1=0$ and $ \, \eta_1=0$, which is twice unstable.
 
  \vspace{0.2cm}
 
  (ii)  A   fixed point, $P_2'$  with ${\mu}'^{*}_2=0,\, {b}'^*_2= {192\pi^2\eps/ 5\,(d_c+4)}$ and $\eta_2'={2\eps/(d_c + 4)}$.  This fixed point has no counterpart  within the two-field model  where $b$ is a function  of $\lambda$ and $\mu$ and, in particular, proportional to $\mu$;  it is  once unstable.  
  
  \vspace{0.2cm}

 (iii)   The infinitely compressible  fixed point  $P_3$   with $\mu^*_3={96\pi^2\eps/(d_c+20)}, \, b^*_3=0$ and  $\eta_3=10\,\epsilon/(20+d_c)$, for which   the bulk modulus $B$  vanishes. It thus identifies with the fixed point $P_3$ of the two-field model;  it  is  once   unstable.  
 
 \vspace{0.2cm}

 (iv)  The  fixed point  $P_4$ with  $\mu^*_4=96\pi^2\,\epsilon/(24+d_c), b^*_4={192\pi^2\eps/5(d_c+24)}$ and $\eta_4=12\,\epsilon/(24+d_c)$ which is  fully stable and controls the flat phase.  It is  located on the stable line $5 b-2\mu=0$ -- corresponding to $(D+1)b-2\mu=0$  in $D$  dimensions -- equivalent  to the line $3\lambda+\mu=0$ in the two-field  model.  It  fully identifies  with  the fixed point $P_4$ of that model. 

 \vspace{0.2cm}

Note finally that, as said above,  in $D=2$,  the tensor $M_{ab,cd}$ vanishes, which is equivalent to the condition $\mu=0$. This implies that  the coordinates of the fixed points all obey this condition.  As a consequence, in $D=2$,  only  one nontrivial  fixed point,  $P_2'$, remains.

\subsubsection{\bf Two-loop order}

At two-loop order, as in the two-field model,  the one-loop  picture is  not radically changed.

 \vspace{0.2cm}

(i)  The Gaussian fixed  point   $P_1$ remains twice  unstable.

 \vspace{0.2cm}

(ii)  At  $P_2'$, ${\mu}'^{*}_2$ still  strictly vanishes whereas   ${b}'^*_2$ is only slightly modified, see Table \ref{tablexpoeff2}.   This  fixed point, as well as  its anomalous dimension  $\eta_2'$ has been first  obtained at two-loop order  by Mauri and Katsnelson \cite{mauri20} in a very recent study of the  Gaussian curvature interaction  (CGI)  model -- see  below.

\begin{table}[htbp]
\begin{center}
\begin{tabular}{|c|l|c|c|c|c|}
\hline
 \rule[-0.15cm]{0cm}{0.5cm} ${\mu}'^{*}_2$ & \hspace{2cm}  $0$      \\
\hline
 \rule[-0.4cm]{0cm}{1.cm}     ${b}'^*_2$  & $\displaystyle\frac{192\pi^2 \, \epsilon}{5(4+d_c)}+\frac{32\pi^2(61 d_c+424)}{75(4+d_c)^3}\, \epsilon^2$          \\
 \hline
\rule[-0.4cm]{0cm}{1.cm}    $\eta_2'$ &  $\displaystyle \frac{2\,\epsilon}{4+d_c}+\frac{d_c(d_c-2)}{6(4+d_c)^3}\,\epsilon^2$        \\
 \hline
\end{tabular}
\end{center}
\caption{Coordinates ${\mu}'^{*}_2$  and ${b}'^*_2$  and  the  corresponding  anomalous dimension $\eta_2'$  of  the   fixed point  $P_2'$  at  order $\epsilon^2$ obtained  from  the  effective model;   $P_2'$  has been first obtained in  \cite{mauri20}.}
\label{tablexpoeff2}
\end{table}

\vspace{0.2cm}

(iii)   The fixed point  $P_3$  is interesting as it has a direct counterpart in the two-field  model, which  allows to study the modifications induced by the change in  model.  Its   coordinates, see Table \ref{tablexpoeff3},   differ from those of the two-field model, see Table  \ref{tablexpo3},    in particular as they  still obey the condition $b_3^*=0$ -- or  $B=0$ -- that puts  $P_3$  just on  the boundary of  the stability region of the theory.   This fact  is an indication that,  within  the two-loop approach of  the two-field model,   the location of  the fixed point  $P_3$  out of the stability region is very likely an artifact of the model or  of its perturbative approach. This  could  also  be a  drawback   of the dimensional regularization  that seems to mismanage $D$-dependent quantities  such as the hypersurface $B=0$.  {\sl Nevertheless}  the  anomalous dimension $\eta_3$, see Table \ref{tablexpoeff3},     coincides exactly with  the two-field result, see Table   \ref{tablexpo3}, which    is a strong check of our computations.  

\begin{table}[htbp]
\begin{center}
\begin{tabular}{|c|l|c|c|c|c|}
\hline
  \rule[-0.4cm]{0cm}{1.cm}   $\mu^*_3$ &  $\displaystyle\frac{96\pi^2 \, \epsilon}{20+d_c}-\frac{80\pi^2(13 d_c+8)}{3(20+d_c)^3}\, \epsilon^2$    \\
\hline
\rule[-0.15cm]{0cm}{0.5cm} $b^*_3$ &  \hspace{1.8cm} $0$          \\
 \hline
\rule[-0.4cm]{0cm}{1.cm}  $\eta_3$ &  $\displaystyle \frac{10\,\epsilon}{20+d_c}-\frac{d_c(37 d_c+950)}{6(20+d_c)^3}\,\epsilon^2$        \\
 \hline
\end{tabular}
\end{center}
\caption{Coordinates $\mu^*_3$ and $b^*_3$ of   the  fixed point  $P_3$  and  the corresponding  anomalous dimension $\eta_3$ at order $\epsilon^2$ obtained  from  the  effective model.}
\label{tablexpoeff3}
\end{table}

(iv)  Finally the fixed point  $P_4$ remains stable and controls the flat phase.  Its coordinates and associated exponent $\eta_4$  are given in Table \ref{tablexpoeff4}.  In the same way as for the fixed point $P_3$,  the coordinates  of $P_4$ at two-loop  order differ  from those obtained from  the two-field model, see Table \ref{tablexpo4}. {\it Also},  these coordinates {\it do  not}   obey the condition $(D+1)b^*_4-2\mu^*_4=(5-\eps)b^*_4-2\mu^*_4=0$ corresponding to the one-loop stability line.  {\sl Nevertheless}, again  the anomalous dimension $\eta_4$ coincides  exactly with the two-field  model  result, see Table \ref{tablexpo4}.

 \begin{table}[htbp]
\begin{center}
\begin{tabular}{|c|l|c|c|c|c|}
\hline
 \rule[-0.4cm]{0cm}{1.cm}  $\mu^*_4$ &  $\displaystyle\frac{96\pi^2 \, \epsilon}{24+d_c}-\frac{32\pi^2(77 d_c+948)}{5(24+d_c)^3}\, \epsilon^2$    \\
\hline
  \rule[-0.4cm]{0cm}{1.cm}  $b^*_4$ & $\displaystyle \frac{192\pi^2 \,\epsilon}{5(24+d_c)}+\frac{64\pi^2(121 d_c+3804)}{25(24+d_c)^3}\, \epsilon^2$          \\
 \hline
 \rule[-0.4cm]{0cm}{1.cm}  $\eta_4$ &  $\displaystyle \frac{12\,\epsilon}{24+d_c}-\frac{6 d_c(d_c+29)}{(24+d_c)^3}\,\epsilon^2$        \\
 \hline
\end{tabular}
\end{center}
\caption{Coordinates $\mu^*_4$ and $b^*_4$ of the flat phase fixed point  $P_4$  and the  corresponding  anomalous dimension $\eta_4$ at order $\epsilon^2$ obtained from   the  effective model.}
\label{tablexpoeff4}
\end{table}

\section{Comparison with previous approaches}

We now discuss  our results compared to the other  techniques -- or other models --  that have been used to investigate the flat phase of membranes. 

\vspace{0.2cm}

{\it SCSA.} The SCSA  has been  studied early \cite{ledoussal92} to investigate the properties  of membranes in any dimension $D$.  It is generally employed using the effective  action (\ref{actioneff})  which is  more  suitable  than (\ref{action})  to establish self-consistent equations.    By construction,  this approach  is one-loop exact. It is also exact at first order in $1/d_c$ and, finally, at $d_c=0$.  Even more remarkably, comparing the anomalous dimensions $\eta_2'$,  $\eta_3$ and $\eta_4$ obtained in this context  to the two-loop results,  see Table  \ref{tablexpoC3},  one observes that the first one is {\it exact} at order $\epsilon^2$ whereas   the latter ones  are  almost  exact at this order as only the coefficients   in $\epsilon^2/d_c^2$ differ   slightly from those of our exact results.  

  There are two important features  of the SCSA  approach that should be underlined.  First,   the solution with a vanishing bare modulus $b=0$, thus corresponding to the fixed point $P_3$,  leads to a vanishing  long-distance effective modulus $b({\bf q})=0$  \cite{ledoussal18}, in agreement with our results $b^*_3=0$.  Second,  under  the conditions  fulfilled to reach the scaling behaviour associated with the  fixed point $P_4$,   one observes   the asymptotic infrared behaviour  \cite{ledoussal92,ledoussal18}: 
 \begin{equation} 
 \displaystyle{\lambda({\bf q})\over \mu({\bf q})}\underset{{\bf q} \to {\bf 0}}{\sim}-{2\over D+2}
 \label{asym}
 \end{equation}
in any dimension $D$ -- which is equivalent to the condition $(D+2)\lambda+2\mu=0$ or, equivalently,  $(D+1)b-2\mu=0$ discussed above. This property has been  proposed to work at all orders of the SCSA and even to be  exact  \cite{gazit09} which  leads  us to wonder about the genuine location of the fixed point $P_4$ found perturbatively at two-loop order  that violates   condition  (\ref{asym}).  

 We finally  recall that, in  $D=2$, one gets, at leading order, $\eta_{SCSA}^{D=2, l}=0.821$ \cite{ledoussal92,ledoussal18}   and, at next-to-leading order,  $\eta_{SCSA}^{D=2, nl}=0.789$ \cite{gazit09}      which  is   inside the range of values  given above and close to  some of the   most recent  results obtained by means of numerical computations (see, e.g., \cite{troster15} that provides $\eta\simeq 0.79$.). 

\begin{widetext}

\begin{table}[h]
\begin{center}
\begin{tabular}{|c|l|c|c|c|c|}
\hline
     &  \hspace{0.5cm}  Two-loop expansion    &  SCSA &  NPRG      \\
       \hline
  \rule[-0.4cm]{0cm}{1.cm}  $\eta_2'$         & $\displaystyle \frac{2\,\epsilon}{4+d_c}+\frac{d_c(d_c-2)}{6(4+d_c)^3}\,\epsilon^2$  &  $\displaystyle \frac{2\,\epsilon}{4+d_c}+\frac{d_c(d_c-2)}{6(4+d_c)^3}\,\epsilon^2$    &       $\displaystyle {2\eps  \over  4+d_c}+ {d_c(10+3d_c)\over 12(4+d_c)^3}\eps^2 $             \\
 \hline
  \rule[-0.4cm]{0cm}{1.cm}  $\eta_3$        & $\displaystyle\ \frac{10\,\epsilon}{20+d_c}-\frac{d_c(37d_c+950)}{6(20+d_c)^3}\,\epsilon^2$ &    $\displaystyle \frac{10\,\epsilon}{20+d_c}-\frac{d_c (37d_c+890)}{6(20+d_c)^3}\,\epsilon^2$         &     $\displaystyle \frac{10\,\epsilon}{20+d_c}-\frac{d_c(69d_c+1430)}{12(20+d_c)^3}\epsilon^2$          \\
  \hline
  \rule[-0.4cm]{0cm}{1.cm}  $\eta_4$         & $\displaystyle\ \frac{12\,\epsilon}{24+d_c}-\frac{6 d_c(d_c+29)}{(24+d_c)^3}\,\epsilon^2$   &    $\displaystyle \frac{12\,\epsilon}{24+d_c}-\frac{6 d_c (d_c+30)}{(24+d_c)^3}\,\epsilon^2$    &  $\displaystyle\frac{12\,\epsilon}{24+d_c} -\frac{\,d_c(11d_c+276)}{2(24+d_c)^3}\,\epsilon^2$                  \\
 \hline
\end{tabular}
\end{center}
\caption{Anomalous dimensions   $\eta_2'$,    $\eta_3$  and  $\eta_4$ obtained  from  the two-loop expansion of either the two-field or the effective model (this paper) -- column  1 --  from  the SCSA \cite{ledoussal92,ledoussal18} -- column 2 -- and from  the NPRG  \cite{kownacki09} --  column 3. The two-loop value of $\eta_2'$ has been first obtained by \cite{mauri20}. }
\label{tablexpoC3}
\end{table}

\end{widetext}

\vspace{0.2cm}

{\it NPRG.}  This  approach  is, as the SCSA,   nonperturbative in the dimensional parameter $\epsilon=4-D$. It is  based  on the use of an exact  RG equation that controls the evolution of a  modified, running  effective action   with the  running scale  \cite{wetterich93c}  (see \cite{bagnuls01,berges02,delamotte03,pawlowski07,rosten12,delamotte12}  for reviews).   Approximations  of this equation  are needed and consist   in  truncating  the running effective action in powers  of the  field-derivatives  (and, if necessary, of the field itself). They however  lead  to RG equations that remain nonperturbative both in $\epsilon$ and in $1/d_c$.   Such a procedure, called derivative  expansion,   has  been  validated   empirically  at order 4 in the derivative of the field \cite{canet03b,depolsi20}  and, more recently,  up to order 6 \cite{balog19b},     since one  observes  a rapid convergence of the  physical quantities with the order in  derivative.  More formal argument for the {\sl convergence}  of the series  --  in contrast  to the  asymptotic nature of the usual, perturbative, series --   have also been given in  \cite{balog19b}.   One should  have in mind  that  this  approach, although  nonperturbative  and, as the SCSA,   exact  in a whole domain of parameters -- at leading  order in $\epsilon$,  in $1/d_c$,  in the  coupling constant controlling the interaction near the lower-critical dimension, at $d_c=0$  --  is  nevertheless   not exact and   generally misses  the next-to-leading order of  the   perturbative approaches.  For instance,   reproducing {\it exactly}   the weak-coupling expansion   at two-loop  order  requires   the knowledge of the infinite series in derivatives \cite{papenbrock95,morris99}.  Yet,  for a given field theory, the ability of the NPRG  to  reproduce satisfactorily this  subleading contribution is a very  good  indication of  its efficiency.   The NPRG    equations for the flat phase of membranes  have been derived at the first order in derivative expansion  in \cite{kownacki09}  and then with help of ansatz  involving the full derivative content in  \cite{braghin10,hasselmann11}. We give  in Table \ref{tablexpoC3},   column 3,   the anomalous dimensions   obtained within this approach  \cite{kownacki09}  and   re-expanded here at second order  in  $\epsilon$.   First,  one notes  that, as in the SCSA case,  the leading order  result is exactly reproduced. Then one can observe that the next-to-leading order  is also numerically close or very close to those obtained within the  two-loop computation.  

It is also interesting to mention that,  for  the  SCSA, the coordinates of the fixed  point $P_3$  obey the condition  of  vanishing bulk modulus 
\begin{equation} 
B=O(\eps^3) 
\label{bnul}
\end{equation}
whereas  those   of the fixed point  $P_4$   obey  the identity:
\begin{equation} 
(6-\eps)\lambda_4^*+2\mu_4^*=O(\eps^3) \ . 
\label{inv}
\end{equation}
The properties (\ref{bnul}) and   (\ref{inv})  are, in fact,  true  nonperturbatively in $\epsilon$ at least within  the first order in the derivative expansion performed in \cite{kownacki09} and, again,   in agreement with the SCSA result (\ref{asym}).

Finally,   one  should recall that  the result obtained in $D=2$ by means of the NPRG approach  \cite{kownacki09,coquand20} $\eta_{NPRG}^{D=2}=0.849(3)$  is also very close to that provided  by several numerical approaches  (see, e.g., \cite{los09,wei14,hasik18} that lead to $\eta\simeq 0.85$).

\vspace{0.2cm}

{\it GCI model.}  We conclude by quoting  a  very recent  -- and  first -- two-loop, weak-coupling  perturbative approach to membranes that has been performed by Mauri and Katsnelson \cite{mauri20}  on a variant of the  effective model (\ref{actioneff})   named Gaussian curvature interaction (GCI)  model.  It is obtained by generalizing to any dimension  $D$ the  simplified   form of the usual effective model (\ref{actioneff}), i.e.\ with $M_{ab,cd}=0$, valid in the particular case $D=2$.  As a consequence the authors of  \cite{mauri20}  get  a -- unique -- nontrivial fixed point which, in our context,  is nothing but the   fixed point  $P_2'$. One of the main results  of their analysis is that the two-loop anomalous dimension $\eta_2'$   coincides  exactly with the corresponding SCSA result,  a fact which is  also  observed in Table \ref{tablexpoC3}.  Our analysis of the complete theory shows  that,  for the stable fixed point $P_4$, a  small discrepancy between  the two-loop and the SCSA results  occurs.

\section{Conclusion}

 We  have performed the two-loop, weak  coupling  analysis of the two models  describing   the flat phase of polymerized  membranes. We have determined the RG equations and the  anomalous dimensions   at this order. We have identified the fixed points,  analyzed  their  properties and computed  the  corresponding  anomalous dimensions.  First,  one notes  that although   the coordinates of the fixed points, as well as several $D$-dependent quantities,   vary from one model to the other,  the anomalous dimensions  at the fixed points   are   very robust  as we get the same values from  the two models.    This provides a very strong check of our computations.  It remains nevertheless to understand  more profoundly  the interplay between the  dimensional regularization used  here and these $D$-dependent quantities that are inherent in theories with space-time symmetries,  such as  the present one.  Second, the very good agreement  between  the anomalous dimensions computed in our  paper   with  those obtained from  the  SCSA and NPRG approaches  is a confirmation of the extreme efficiency  of these  last methods in the context of  the theory of the flat phase of polymerized membranes.    As said,  these two approaches have in common that they both reproduce  exactly --  by construction --  the  leading order  of all   usual perturbative approaches.   This, however,  does  not   explain   their  singular  achievements  here  which  more likely  rely  on the very nature of the flat phase of membranes itself.   This is under investigation.

\acknowledgements

We wish to thank warmly  J.  Gracey, M. Kompaniets and  K. J. Wiese for  very fruitful discussions.


\begin{thebibliography}{77}
\expandafter\ifx\csname natexlab\endcsname\relax\def\natexlab#1{#1}\fi
\expandafter\ifx\csname bibnamefont\endcsname\relax
  \def\bibnamefont#1{#1}\fi
\expandafter\ifx\csname bibfnamefont\endcsname\relax
  \def\bibfnamefont#1{#1}\fi
\expandafter\ifx\csname citenamefont\endcsname\relax
  \def\citenamefont#1{#1}\fi
\expandafter\ifx\csname url\endcsname\relax
  \def\url#1{\texttt{#1}}\fi
\expandafter\ifx\csname urlprefix\endcsname\relax\def\urlprefix{URL }\fi
\providecommand{\bibinfo}[2]{#2}
\providecommand{\eprint}[2][]{\url{#2}}

\bibitem[{\citenamefont{Nelson et~al.}(2004)\citenamefont{Nelson, Piran, and
  Weinberg}}]{proceedings89}
\bibinfo{editor}{\bibfnamefont{D.~R.} \bibnamefont{Nelson}},
  \bibinfo{editor}{\bibfnamefont{T.}~\bibnamefont{Piran}}, \bibnamefont{and}
  \bibinfo{editor}{\bibfnamefont{S.}~\bibnamefont{Weinberg}}, eds.,
  \emph{\bibinfo{title}{Proceedings of the Fifth Jerusalem Winter School for
  Theoretical Physics}} (\bibinfo{publisher}{World Scientific, Singapore},
  \bibinfo{year}{2004}), \bibinfo{edition}{2nd} ed.

\bibitem[{\citenamefont{Bowick and Travesset}(2001)}]{bowick01}
\bibinfo{author}{\bibfnamefont{M.~J.} \bibnamefont{Bowick}} \bibnamefont{and}
  \bibinfo{author}{\bibfnamefont{A.}~\bibnamefont{Travesset}},
  \bibinfo{journal}{Phys. Rep.} \textbf{\bibinfo{volume}{344}},
  \bibinfo{pages}{255} (\bibinfo{year}{2001}).

\bibitem[{\citenamefont{Polyakov}(1987)}]{polyakov87}
\bibinfo{author}{\bibfnamefont{A.~M.} \bibnamefont{Polyakov}},
  \emph{\bibinfo{title}{Gauge Field and Strings}} (\bibinfo{publisher}{Gordon
  and Breach}, \bibinfo{year}{1987}).

\bibitem[{\citenamefont{David}(1989)}]{david89}
\bibinfo{author}{\bibfnamefont{F.}~\bibnamefont{David}},
  \bibinfo{journal}{Phys. Rep.} \textbf{\bibinfo{volume}{184}},
  \bibinfo{pages}{221} (\bibinfo{year}{1989}).

\bibitem[{\citenamefont{Wheater}(1994)}]{wheater94}
\bibinfo{author}{\bibfnamefont{J.}~\bibnamefont{Wheater}}, \bibinfo{journal}{J.
  Phys. A} \textbf{\bibinfo{volume}{27}}, \bibinfo{pages}{3323}
  (\bibinfo{year}{1994}).

\bibitem[{\citenamefont{{{See the contribution of F. David in
  [1]}}}()}]{david04}
\bibinfo{author}{\bibnamefont{{{See the contribution of F. David in [1]}}}}.

\bibitem[{\citenamefont{Schmidt et~al.}(1993)\citenamefont{Schmidt, Svoboda,
  Lei, Petsche, Berman, Safinya, and Grest}}]{schmidt93}
\bibinfo{author}{\bibfnamefont{C.~F.} \bibnamefont{Schmidt}},
  \bibinfo{author}{\bibfnamefont{K.}~\bibnamefont{Svoboda}},
  \bibinfo{author}{\bibfnamefont{N.}~\bibnamefont{Lei}},
  \bibinfo{author}{\bibfnamefont{I.~B.} \bibnamefont{Petsche}},
  \bibinfo{author}{\bibfnamefont{L.~E.} \bibnamefont{Berman}},
  \bibinfo{author}{\bibfnamefont{C.~R.} \bibnamefont{Safinya}},
  \bibnamefont{and} \bibinfo{author}{\bibfnamefont{G.~S.} \bibnamefont{Grest}},
  \bibinfo{journal}{Science} \textbf{\bibinfo{volume}{259}},
  \bibinfo{pages}{952} (\bibinfo{year}{1993}).

\bibitem[{\citenamefont{Novoselov et~al.}(2004)\citenamefont{Novoselov, Geim,
  Morozov, Jiang, Zhang, Dubonos, Gregorieva, and Firsov}}]{novoselov04}
\bibinfo{author}{\bibfnamefont{K.~S.} \bibnamefont{Novoselov}},
  \bibinfo{author}{\bibfnamefont{A.~K.} \bibnamefont{Geim}},
  \bibinfo{author}{\bibfnamefont{S.~V.} \bibnamefont{Morozov}},
  \bibinfo{author}{\bibfnamefont{D.}~\bibnamefont{Jiang}},
  \bibinfo{author}{\bibfnamefont{Y.}~\bibnamefont{Zhang}},
  \bibinfo{author}{\bibfnamefont{S.~V.} \bibnamefont{Dubonos}},
  \bibinfo{author}{\bibfnamefont{I.~V.} \bibnamefont{Gregorieva}},
  \bibnamefont{and} \bibinfo{author}{\bibfnamefont{A.~A.}
  \bibnamefont{Firsov}}, \bibinfo{journal}{Science}
  \textbf{\bibinfo{volume}{306}}, \bibinfo{pages}{666} (\bibinfo{year}{2004}).

\bibitem[{\citenamefont{Novoselov et~al.}(2005)\citenamefont{Novoselov, Geim,
  Morozov, Jiang, Katsnelson, Gregorieva, Dubonos, and Firsov}}]{novoselov05}
\bibinfo{author}{\bibfnamefont{K.~S.} \bibnamefont{Novoselov}},
  \bibinfo{author}{\bibfnamefont{A.~K.} \bibnamefont{Geim}},
  \bibinfo{author}{\bibfnamefont{S.~V.} \bibnamefont{Morozov}},
  \bibinfo{author}{\bibfnamefont{D.}~\bibnamefont{Jiang}},
  \bibinfo{author}{\bibfnamefont{M.~I.} \bibnamefont{Katsnelson}},
  \bibinfo{author}{\bibfnamefont{I.~V.} \bibnamefont{Gregorieva}},
  \bibinfo{author}{\bibfnamefont{S.~V.} \bibnamefont{Dubonos}},
  \bibnamefont{and} \bibinfo{author}{\bibfnamefont{A.~A.}
  \bibnamefont{Firsov}}, \bibinfo{journal}{Nature}
  \textbf{\bibinfo{volume}{438}}, \bibinfo{pages}{197} (\bibinfo{year}{2005}).

\bibitem[{\citenamefont{Katsnelson}(2012)}]{katsnelson12}
\bibinfo{author}{\bibfnamefont{M.~I.} \bibnamefont{Katsnelson}},
  \emph{\bibinfo{title}{Graphene: Carbon in Two Dimensions}}
  (\bibinfo{publisher}{Cambridge University Press},
  \bibinfo{address}{{Cambridge, U.K.}}, \bibinfo{year}{2012}).

\bibitem[{\citenamefont{{{See the contribution of D. R. Nelson in
  Ref.[1]}}}()}]{nelson04}
\bibinfo{author}{\bibnamefont{{{See the contribution of D. R. Nelson in
  Ref.[1]}}}}.

\bibitem[{\citenamefont{{D. R. Nelson}}(2002)}]{nelson02}
\bibinfo{author}{\bibnamefont{{D. R. Nelson}}}, \emph{\bibinfo{title}{Defects
  and Geometry in Condensed Matter Physics}} (\bibinfo{publisher}{Cambridge
  University Press, Cambridge, UK}, \bibinfo{year}{2002}).

\bibitem[{\citenamefont{de~Gennes and Taupin}(1982)}]{degennes82}
\bibinfo{author}{\bibfnamefont{P.-G.} \bibnamefont{de~Gennes}}
  \bibnamefont{and} \bibinfo{author}{\bibfnamefont{C.}~\bibnamefont{Taupin}},
  \bibinfo{journal}{J. Phys. Chem} \textbf{\bibinfo{volume}{86}},
  \bibinfo{pages}{2294} (\bibinfo{year}{1982}).

\bibitem[{\citenamefont{Peliti and Leibler}(1985)}]{peliti85}
\bibinfo{author}{\bibfnamefont{L.}~\bibnamefont{Peliti}} \bibnamefont{and}
  \bibinfo{author}{\bibfnamefont{S.}~\bibnamefont{Leibler}},
  \bibinfo{journal}{Phys. Rev. Lett.} \textbf{\bibinfo{volume}{54}},
  \bibinfo{pages}{1690} (\bibinfo{year}{1985}).

\bibitem[{\citenamefont{Helfrich}(1985)}]{helfrich85}
\bibinfo{author}{\bibfnamefont{W.}~\bibnamefont{Helfrich}},
  \bibinfo{journal}{J. Phys. France} \textbf{\bibinfo{volume}{46}},
  \bibinfo{pages}{1263} (\bibinfo{year}{1985}).

\bibitem[{\citenamefont{Mermin and Wagner}(1966)}]{mermin66}
\bibinfo{author}{\bibfnamefont{N.~D.} \bibnamefont{Mermin}} \bibnamefont{and}
  \bibinfo{author}{\bibfnamefont{H.}~\bibnamefont{Wagner}},
  \bibinfo{journal}{Phys. Rev. Lett.} \textbf{\bibinfo{volume}{17}},
  \bibinfo{pages}{1133} (\bibinfo{year}{1966}).

\bibitem[{\citenamefont{Nelson and Peliti}(1987)}]{nelson87}
\bibinfo{author}{\bibfnamefont{D.~R.} \bibnamefont{Nelson}} \bibnamefont{and}
  \bibinfo{author}{\bibfnamefont{L.}~\bibnamefont{Peliti}},
  \bibinfo{journal}{J. Phys. (Paris)} \textbf{\bibinfo{volume}{48}},
  \bibinfo{pages}{1085} (\bibinfo{year}{1987}).

\bibitem[{\citenamefont{Aronovitz and Lubensky}(1988)}]{aronovitz88}
\bibinfo{author}{\bibfnamefont{J.~A.} \bibnamefont{Aronovitz}}
  \bibnamefont{and} \bibinfo{author}{\bibfnamefont{T.~C.}
  \bibnamefont{Lubensky}}, \bibinfo{journal}{Phys. Rev. Lett.}
  \textbf{\bibinfo{volume}{60}}, \bibinfo{pages}{2634} (\bibinfo{year}{1988}).

\bibitem[{\citenamefont{David and Guitter}(1988)}]{david88}
\bibinfo{author}{\bibfnamefont{F.}~\bibnamefont{David}} \bibnamefont{and}
  \bibinfo{author}{\bibfnamefont{E.}~\bibnamefont{Guitter}},
  \bibinfo{journal}{Europhys. Lett.} \textbf{\bibinfo{volume}{5}},
  \bibinfo{pages}{709} (\bibinfo{year}{1988}).

\bibitem[{\citenamefont{Guitter et~al.}(1989)\citenamefont{Guitter, David,
  Leibler, and Peliti}}]{guitter89}
\bibinfo{author}{\bibfnamefont{E.}~\bibnamefont{Guitter}},
  \bibinfo{author}{\bibfnamefont{F.}~\bibnamefont{David}},
  \bibinfo{author}{\bibfnamefont{S.}~\bibnamefont{Leibler}}, \bibnamefont{and}
  \bibinfo{author}{\bibfnamefont{L.}~\bibnamefont{Peliti}},
  \bibinfo{journal}{J. Phys. (Paris)} \textbf{\bibinfo{volume}{50}},
  \bibinfo{pages}{1787} (\bibinfo{year}{1989}).

\bibitem[{\citenamefont{{M. Paczuski and M. Kardar and D. R.
  Nelson}}(1988)}]{paczuski88}
\bibinfo{author}{\bibnamefont{{M. Paczuski and M. Kardar and D. R. Nelson}}},
  \bibinfo{journal}{Phys. Rev. Lett.} \textbf{\bibinfo{volume}{60}},
  \bibinfo{pages}{2638} (\bibinfo{year}{1988}).

\bibitem[{\citenamefont{Aronovitz et~al.}(1989)\citenamefont{Aronovitz,
  Golubovic, and Lubensky}}]{aronovitz89}
\bibinfo{author}{\bibfnamefont{J.~A.} \bibnamefont{Aronovitz}},
  \bibinfo{author}{\bibfnamefont{L.}~\bibnamefont{Golubovic}},
  \bibnamefont{and} \bibinfo{author}{\bibfnamefont{T.~C.}
  \bibnamefont{Lubensky}}, \bibinfo{journal}{J. Phys. (Paris)}
  \textbf{\bibinfo{volume}{50}}, \bibinfo{pages}{609} (\bibinfo{year}{1989}).

\bibitem[{\citenamefont{{{J.-P. } Kownacki} and Diep}(2002)}]{kownacki02}
\bibinfo{author}{\bibnamefont{{{J.-P. } Kownacki}}} \bibnamefont{and}
  \bibinfo{author}{\bibfnamefont{H.}~\bibnamefont{Diep}},
  \bibinfo{journal}{Phys. Rev. E} \textbf{\bibinfo{volume}{66}},
  \bibinfo{pages}{066105} (\bibinfo{year}{2002}).

\bibitem[{\citenamefont{{{H. Koibuchi and N. Kusano and A. Nidaira and K.
  Suzuki and M. Yamada}}}(2004)}]{koibuchi04}
\bibinfo{author}{\bibnamefont{{{H. Koibuchi and N. Kusano and A. Nidaira and K.
  Suzuki and M. Yamada}}}}, \bibinfo{journal}{Phys. Rev. E}
  \textbf{\bibinfo{volume}{69}}, \bibinfo{pages}{066139}
  (\bibinfo{year}{2004}).

\bibitem[{\citenamefont{{{H. Koibuchi and T. Kuwahata}}}(2005)}]{koibuchi05}
\bibinfo{author}{\bibnamefont{{{H. Koibuchi and T. Kuwahata}}}},
  \bibinfo{journal}{Phys. Rev. E} \textbf{\bibinfo{volume}{72}},
  \bibinfo{pages}{026124} (\bibinfo{year}{2005}).

\bibitem[{\citenamefont{{{J.-P. } Kownacki} and Mouhanna}(2009)}]{kownacki09}
\bibinfo{author}{\bibnamefont{{{J.-P. } Kownacki}}} \bibnamefont{and}
  \bibinfo{author}{\bibfnamefont{D.}~\bibnamefont{Mouhanna}},
  \bibinfo{journal}{Phys. Rev. E} \textbf{\bibinfo{volume}{79}},
  \bibinfo{pages}{040101(R)} (\bibinfo{year}{2009}).

\bibitem[{\citenamefont{{{H. Koibuchi and A. Shobukhov}}}(2014)}]{koibuchi14}
\bibinfo{author}{\bibnamefont{{{H. Koibuchi and A. Shobukhov}}}},
  \bibinfo{journal}{Int. J. Mod. Phys. C} \textbf{\bibinfo{volume}{25}},
  \bibinfo{pages}{1450033} (\bibinfo{year}{2014}).

\bibitem[{\citenamefont{Essafi et~al.}(2014)\citenamefont{Essafi, Kownacki, and
  Mouhanna}}]{essafi14}
\bibinfo{author}{\bibfnamefont{K.}~\bibnamefont{Essafi}},
  \bibinfo{author}{\bibfnamefont{J.-P.} \bibnamefont{Kownacki}},
  \bibnamefont{and} \bibinfo{author}{\bibfnamefont{D.}~\bibnamefont{Mouhanna}},
  \bibinfo{journal}{Phys. Rev. E} \textbf{\bibinfo{volume}{89}},
  \bibinfo{pages}{042101} (\bibinfo{year}{2014}).

\bibitem[{\citenamefont{{U. Satoshi and H. Koibuchi}}(2016)}]{satoshi16}
\bibinfo{author}{\bibnamefont{{U. Satoshi and H. Koibuchi}}},
  \bibinfo{journal}{J. Stat. Phys.} \textbf{\bibinfo{volume}{162}},
  \bibinfo{pages}{701} (\bibinfo{year}{2016}).

\bibitem[{\citenamefont{{R. Cuerno, R. Gallardo Caballero, A.
  Gordillo-Guerrero, P. Monroy and J. J. Ruiz-Lorenzo}}(2016)}]{cuerno16}
\bibinfo{author}{\bibnamefont{{R. Cuerno, R. Gallardo Caballero, A.
  Gordillo-Guerrero, P. Monroy and J. J. Ruiz-Lorenzo}}},
  \bibinfo{journal}{Phys. Rev. E} \textbf{\bibinfo{volume}{93}},
  \bibinfo{pages}{022111} (\bibinfo{year}{2016}).

\bibitem[{\citenamefont{Bowick et~al.}(2001)\citenamefont{Bowick, Cacciuto,
  Thoroleifsson, and Travesset}}]{bowick01b}
\bibinfo{author}{\bibfnamefont{M.~J.} \bibnamefont{Bowick}},
  \bibinfo{author}{\bibfnamefont{A.}~\bibnamefont{Cacciuto}},
  \bibinfo{author}{\bibfnamefont{G.}~\bibnamefont{Thoroleifsson}},
  \bibnamefont{and}
  \bibinfo{author}{\bibfnamefont{A.}~\bibnamefont{Travesset}},
  \bibinfo{journal}{Eur. Phys. J. E} \textbf{\bibinfo{volume}{{5}}},
  \bibinfo{pages}{149} (\bibinfo{year}{2001}).

\bibitem[{\citenamefont{Guitter et~al.}(1988)\citenamefont{Guitter, David,
  Leibler, and Peliti}}]{guitter88}
\bibinfo{author}{\bibfnamefont{E.}~\bibnamefont{Guitter}},
  \bibinfo{author}{\bibfnamefont{F.}~\bibnamefont{David}},
  \bibinfo{author}{\bibfnamefont{S.}~\bibnamefont{Leibler}}, \bibnamefont{and}
  \bibinfo{author}{\bibfnamefont{L.}~\bibnamefont{Peliti}},
  \bibinfo{journal}{Phys. Rev. Lett.} \textbf{\bibinfo{volume}{61}},
  \bibinfo{pages}{2949} (\bibinfo{year}{1988}).

\bibitem[{\citenamefont{{I. V. Gornyi, V. Yu. Kachorovskii and A. D.
  Mirlin}}(2015)}]{gornyi15}
\bibinfo{author}{\bibnamefont{{I. V. Gornyi, V. Yu. Kachorovskii and A. D.
  Mirlin}}}, \bibinfo{journal}{Phys. Rev. B} \textbf{\bibinfo{volume}{92}},
  \bibinfo{pages}{155428} (\bibinfo{year}{2015}).

\bibitem[{\citenamefont{{Le Doussal} and Radzihovsky}(1992)}]{ledoussal92}
\bibinfo{author}{\bibfnamefont{P.}~\bibnamefont{{Le Doussal}}}
  \bibnamefont{and}
  \bibinfo{author}{\bibfnamefont{L.}~\bibnamefont{Radzihovsky}},
  \bibinfo{journal}{Phys. Rev. Lett.} \textbf{\bibinfo{volume}{69}},
  \bibinfo{pages}{1209} (\bibinfo{year}{1992}).

\bibitem[{\citenamefont{{P. Le Doussal and L.
  Radzihovsky}}(2018)}]{ledoussal18}
\bibinfo{author}{\bibnamefont{{P. Le Doussal and L. Radzihovsky}}},
  \bibinfo{journal}{Ann. Phys. (N.Y.)} \textbf{\bibinfo{volume}{392}},
  \bibinfo{pages}{340} (\bibinfo{year}{2018}).

\bibitem[{\citenamefont{Coquand}(2019)}]{coquand19b}
\bibinfo{author}{\bibfnamefont{O.}~\bibnamefont{Coquand}},
  \bibinfo{journal}{Phys. Rev B} \textbf{\bibinfo{volume}{100}},
  \bibinfo{pages}{125406} (\bibinfo{year}{2019}).

\bibitem[{\citenamefont{Coquand
  et~al.}(2020{\natexlab{a}})\citenamefont{Coquand, Mouhanna, and
  Teber}}]{coquand20b}
\bibinfo{author}{\bibfnamefont{O.}~\bibnamefont{Coquand}},
  \bibinfo{author}{\bibfnamefont{D.}~\bibnamefont{Mouhanna}}, \bibnamefont{and}
  \bibinfo{author}{\bibfnamefont{S.}~\bibnamefont{Teber}},
  \bibinfo{journal}{unpublished}  (\bibinfo{year}{2020}{\natexlab{a}}).

\bibitem[{\citenamefont{{D. R. Saykin, I. V. Gornyi, V. Yu. Kachorovskii and I.
  S. Burmistrov}}(2020)}]{saykin20}
\bibinfo{author}{\bibnamefont{{D. R. Saykin, I. V. Gornyi, V. Yu. Kachorovskii
  and I. S. Burmistrov}}}, \bibinfo{journal}{Ann. Phys. (N.Y.)}
  \textbf{\bibinfo{volume}{414}}, \bibinfo{pages}{168108}
  (\bibinfo{year}{2020}).

\bibitem[{\citenamefont{{K. V. Zakharchenko, R. Rold\'an, A. Fasolino and M. I.
  Katsnelson}}(2010)}]{zakharchenko10}
\bibinfo{author}{\bibnamefont{{K. V. Zakharchenko, R. Rold\'an, A. Fasolino and
  M. I. Katsnelson}}}, \bibinfo{journal}{Phys. Rev. B}
  \textbf{\bibinfo{volume}{82}}, \bibinfo{pages}{125435}
  (\bibinfo{year}{2010}).

\bibitem[{\citenamefont{{R. Rold\'an, A. Fasolino, K. V. Zakharchenko, and M.
  I. Katsnelson}}(2011)}]{roldan11}
\bibinfo{author}{\bibnamefont{{R. Rold\'an, A. Fasolino, K. V. Zakharchenko,
  and M. I. Katsnelson}}}, \bibinfo{journal}{Phys. Rev. B}
  \textbf{\bibinfo{volume}{83}}, \bibinfo{pages}{174104}
  (\bibinfo{year}{2011}).

\bibitem[{\citenamefont{Gazit}(2009)}]{gazit09}
\bibinfo{author}{\bibfnamefont{D.}~\bibnamefont{Gazit}},
  \bibinfo{journal}{Phys. Rev. E} \textbf{\bibinfo{volume}{80}},
  \bibinfo{pages}{041117} (\bibinfo{year}{2009}).

\bibitem[{\citenamefont{Essafi et~al.}(2011)\citenamefont{Essafi, Kownacki, and
  Mouhanna}}]{essafi11}
\bibinfo{author}{\bibfnamefont{K.}~\bibnamefont{Essafi}},
  \bibinfo{author}{\bibfnamefont{J.-P.} \bibnamefont{Kownacki}},
  \bibnamefont{and} \bibinfo{author}{\bibfnamefont{D.}~\bibnamefont{Mouhanna}},
  \bibinfo{journal}{Phys. Rev. Lett.} \textbf{\bibinfo{volume}{106}},
  \bibinfo{pages}{128102} (\bibinfo{year}{2011}).

\bibitem[{\citenamefont{Coquand and Mouhanna}(2016)}]{coquand16a}
\bibinfo{author}{\bibfnamefont{O.}~\bibnamefont{Coquand}} \bibnamefont{and}
  \bibinfo{author}{\bibfnamefont{D.}~\bibnamefont{Mouhanna}},
  \bibinfo{journal}{Phys. Rev. E} \textbf{\bibinfo{volume}{94}},
  \bibinfo{pages}{032125} (\bibinfo{year}{2016}).

\bibitem[{\citenamefont{Coquand et~al.}(2018)\citenamefont{Coquand, Essafi,
  Kownacki, and Mouhanna}}]{coquand18}
\bibinfo{author}{\bibfnamefont{O.}~\bibnamefont{Coquand}},
  \bibinfo{author}{\bibfnamefont{K.}~\bibnamefont{Essafi}},
  \bibinfo{author}{\bibfnamefont{J.-P.} \bibnamefont{Kownacki}},
  \bibnamefont{and} \bibinfo{author}{\bibfnamefont{D.}~\bibnamefont{Mouhanna}},
  \bibinfo{journal}{Phys. Rev E} \textbf{\bibinfo{volume}{97}},
  \bibinfo{pages}{{030102(R)}} (\bibinfo{year}{2018}).

\bibitem[{\citenamefont{Coquand
  et~al.}(2020{\natexlab{b}})\citenamefont{Coquand, Essafi, Kownacki, and
  Mouhanna}}]{coquand20}
\bibinfo{author}{\bibfnamefont{O.}~\bibnamefont{Coquand}},
  \bibinfo{author}{\bibfnamefont{K.}~\bibnamefont{Essafi}},
  \bibinfo{author}{\bibfnamefont{J.-P.} \bibnamefont{Kownacki}},
  \bibnamefont{and} \bibinfo{author}{\bibfnamefont{D.}~\bibnamefont{Mouhanna}},
  \bibinfo{journal}{Phys. Rev. E} \textbf{\bibinfo{volume}{101}},
  \bibinfo{pages}{042602} (\bibinfo{year}{2020}{\natexlab{b}}).

\bibitem[{\citenamefont{{F. L. Braghin and N. Hasselmann}}(2010)}]{braghin10}
\bibinfo{author}{\bibnamefont{{F. L. Braghin and N. Hasselmann}}},
  \bibinfo{journal}{Phys. Rev. B} \textbf{\bibinfo{volume}{82}},
  \bibinfo{pages}{035407} (\bibinfo{year}{2010}).

\bibitem[{\citenamefont{{N. Hasselmann and F. L.
  Braghin}}(2011)}]{hasselmann11}
\bibinfo{author}{\bibnamefont{{N. Hasselmann and F. L. Braghin}}},
  \bibinfo{journal}{Phys. Rev. E} \textbf{\bibinfo{volume}{83}},
  \bibinfo{pages}{031137} (\bibinfo{year}{2011}).

\bibitem[{\citenamefont{{A. Mauri and M.I. Katsnelson}}(2020)}]{mauri20}
\bibinfo{author}{\bibnamefont{{A. Mauri and M.I. Katsnelson}}},
  \bibinfo{journal}{Nucl. Phys. B} \textbf{\bibinfo{volume}{956}},
  \bibinfo{pages}{115040} (\bibinfo{year}{2020}).

\bibitem[{\citenamefont{Smirnov and Chetyrkin}(1984)}]{smirnov83}
\bibinfo{author}{\bibfnamefont{V.~A.} \bibnamefont{Smirnov}} \bibnamefont{and}
  \bibinfo{author}{\bibfnamefont{K.~G.} \bibnamefont{Chetyrkin}},
  \bibinfo{journal}{Theor. Math. Phys.} \textbf{\bibinfo{volume}{56}},
  \bibinfo{pages}{770} (\bibinfo{year}{1984}).

\bibitem[{\citenamefont{Bogoliubov and Parasiuk}(1957)}]{Bogoliubov:1957gp}
\bibinfo{author}{\bibfnamefont{N.~N.} \bibnamefont{Bogoliubov}}
  \bibnamefont{and} \bibinfo{author}{\bibfnamefont{O.~S.}
  \bibnamefont{Parasiuk}}, \bibinfo{journal}{Acta Math.}
  \textbf{\bibinfo{volume}{97}}, \bibinfo{pages}{227} (\bibinfo{year}{1957}).

\bibitem[{\citenamefont{Hepp}(1966)}]{Hepp:1966eg}
\bibinfo{author}{\bibfnamefont{K.}~\bibnamefont{Hepp}},
  \bibinfo{journal}{Commun. Math. Phys.} \textbf{\bibinfo{volume}{2}},
  \bibinfo{pages}{301} (\bibinfo{year}{1966}).

\bibitem[{\citenamefont{Zimmermann}(1969)}]{Zimmermann:1969jj}
\bibinfo{author}{\bibfnamefont{W.}~\bibnamefont{Zimmermann}},
  \bibinfo{journal}{Commun. Math. Phys.} \textbf{\bibinfo{volume}{15}},
  \bibinfo{pages}{208} (\bibinfo{year}{1969}).

\bibitem[{\citenamefont{Lee}(2014)}]{Lee:2013mka}
\bibinfo{author}{\bibfnamefont{R.~N.} \bibnamefont{Lee}}, \bibinfo{journal}{{J.
  Phys.: Conf. Ser.}} \textbf{\bibinfo{volume}{523}}, \bibinfo{pages}{012059}
  (\bibinfo{year}{2014}).

\bibitem[{\citenamefont{Kotikov and Teber}(2019)}]{Kotikov:2018wxe}
\bibinfo{author}{\bibfnamefont{A.~V.} \bibnamefont{Kotikov}} \bibnamefont{and}
  \bibinfo{author}{\bibfnamefont{S.}~\bibnamefont{Teber}},
  \bibinfo{journal}{Phys. Part. Nucl.} \textbf{\bibinfo{volume}{50}},
  \bibinfo{pages}{1} (\bibinfo{year}{2019}).

\bibitem[{\citenamefont{Guitter et~al.}(1990)\citenamefont{Guitter, Leibler,
  Maggs, and David}}]{guitter90}
\bibinfo{author}{\bibfnamefont{E.}~\bibnamefont{Guitter}},
  \bibinfo{author}{\bibfnamefont{S.}~\bibnamefont{Leibler}},
  \bibinfo{author}{\bibfnamefont{A.}~\bibnamefont{Maggs}}, \bibnamefont{and}
  \bibinfo{author}{\bibfnamefont{F.}~\bibnamefont{David}}, \bibinfo{journal}{J.
  Phys. (Paris)} \textbf{\bibinfo{volume}{51}}, \bibinfo{pages}{1055}
  (\bibinfo{year}{1990}).

\bibitem[{\citenamefont{Zhang et~al.}(1993)\citenamefont{Zhang, Davis, and
  Kroll}}]{zhang93}
\bibinfo{author}{\bibfnamefont{Z.}~\bibnamefont{Zhang}},
  \bibinfo{author}{\bibfnamefont{H.~T.} \bibnamefont{Davis}}, \bibnamefont{and}
  \bibinfo{author}{\bibfnamefont{D.~M.} \bibnamefont{Kroll}},
  \bibinfo{journal}{Phys. Rev. E} \textbf{\bibinfo{volume}{48}},
  \bibinfo{pages}{{R651}} (\bibinfo{year}{1993}).

\bibitem[{\citenamefont{{{M. J. Bowick et. al.}}}(1996)}]{bowick96}
\bibinfo{author}{\bibnamefont{{{M. J. Bowick et. al.}}}}, \bibinfo{journal}{J.
  Phys. (France) I} \textbf{\bibinfo{volume}{6}}, \bibinfo{pages}{1321}
  (\bibinfo{year}{1996}).

\bibitem[{\citenamefont{Gompper and Kroll}(1997)}]{gompper97}
\bibinfo{author}{\bibfnamefont{G.}~\bibnamefont{Gompper}} \bibnamefont{and}
  \bibinfo{author}{\bibfnamefont{D.~M.} \bibnamefont{Kroll}},
  \bibinfo{journal}{J. Phys: Condens. Matter} \textbf{\bibinfo{volume}{9}},
  \bibinfo{pages}{8795} (\bibinfo{year}{1997}).

\bibitem[{\citenamefont{{J. H. Los, M. I. Katsnelson, O. V. Yazyev, K. V.
  Zakharchenko, and A. Fasolino}}(2009)}]{los09}
\bibinfo{author}{\bibnamefont{{J. H. Los, M. I. Katsnelson, O. V. Yazyev, K. V.
  Zakharchenko, and A. Fasolino}}}, \bibinfo{journal}{Phys. Rev. B}
  \textbf{\bibinfo{volume}{80}}, \bibinfo{pages}{121405(R)}
  (\bibinfo{year}{2009}).

\bibitem[{\citenamefont{{A. Tröster}}(2013)}]{troster13}
\bibinfo{author}{\bibnamefont{{A. Tröster}}}, \bibinfo{journal}{Phys. Rev. B}
  \textbf{\bibinfo{volume}{87}}, \bibinfo{pages}{104112}
  (\bibinfo{year}{2013}).

\bibitem[{\citenamefont{Wein and Wang}(2014)}]{wei14}
\bibinfo{author}{\bibfnamefont{D.}~\bibnamefont{Wein}} \bibnamefont{and}
  \bibinfo{author}{\bibfnamefont{F.}~\bibnamefont{Wang}}, \bibinfo{journal}{J.
  Chem. Phys.} \textbf{\bibinfo{volume}{141}}, \bibinfo{pages}{144701}
  (\bibinfo{year}{2014}).

\bibitem[{\citenamefont{{A. Tröster}}(2015)}]{troster15}
\bibinfo{author}{\bibnamefont{{A. Tröster}}}, \bibinfo{journal}{Phys. Rev. B}
  \textbf{\bibinfo{volume}{91}}, \bibinfo{pages}{022132}
  (\bibinfo{year}{2015}).

\bibitem[{\citenamefont{{J. H. Los, A. Fasolino and M. I.
  Katsnelson}}(2016)}]{los16}
\bibinfo{author}{\bibnamefont{{J. H. Los, A. Fasolino and M. I. Katsnelson}}},
  \bibinfo{journal}{Phys. Rev. Lett.} \textbf{\bibinfo{volume}{116}},
  \bibinfo{pages}{015901} (\bibinfo{year}{2016}).

\bibitem[{\citenamefont{{A. Kosmrlj and D.R. Nelson}}(2017)}]{kosmrlj17}
\bibinfo{author}{\bibnamefont{{A. Kosmrlj and D.R. Nelson}}},
  \bibinfo{journal}{Phys. Rev. X} \textbf{\bibinfo{volume}{7}},
  \bibinfo{pages}{011002} (\bibinfo{year}{2017}).

\bibitem[{\citenamefont{{J. Hasik, E. Tosatti and R.
  Martonak}}(2018)}]{hasik18}
\bibinfo{author}{\bibnamefont{{J. Hasik, E. Tosatti and R. Martonak}}},
  \bibinfo{journal}{Phys. Rev. B} \textbf{\bibinfo{volume}{97}},
  \bibinfo{pages}{140301(R)} (\bibinfo{year}{2018}).

\bibitem[{\citenamefont{Wetterich}(1993)}]{wetterich93c}
\bibinfo{author}{\bibfnamefont{C.}~\bibnamefont{Wetterich}},
  \bibinfo{journal}{Phys. Lett. B} \textbf{\bibinfo{volume}{301}},
  \bibinfo{pages}{90} (\bibinfo{year}{1993}).

\bibitem[{\citenamefont{Bagnuls and Bervillier}(2001)}]{bagnuls01}
\bibinfo{author}{\bibfnamefont{C.}~\bibnamefont{Bagnuls}} \bibnamefont{and}
  \bibinfo{author}{\bibfnamefont{C.}~\bibnamefont{Bervillier}},
  \bibinfo{journal}{Phys. Rep.} \textbf{\bibinfo{volume}{348}},
  \bibinfo{pages}{91} (\bibinfo{year}{2001}).

\bibitem[{\citenamefont{Berges et~al.}(2002)\citenamefont{Berges, Tetradis, and
  Wetterich}}]{berges02}
\bibinfo{author}{\bibfnamefont{J.}~\bibnamefont{Berges}},
  \bibinfo{author}{\bibfnamefont{N.}~\bibnamefont{Tetradis}}, \bibnamefont{and}
  \bibinfo{author}{\bibfnamefont{C.}~\bibnamefont{Wetterich}},
  \bibinfo{journal}{Phys. Rep.} \textbf{\bibinfo{volume}{363}},
  \bibinfo{pages}{223} (\bibinfo{year}{2002}).

\bibitem[{\citenamefont{Delamotte et~al.}(2004)\citenamefont{Delamotte,
  Mouhanna, and Tissier}}]{delamotte03}
\bibinfo{author}{\bibfnamefont{B.}~\bibnamefont{Delamotte}},
  \bibinfo{author}{\bibfnamefont{D.}~\bibnamefont{Mouhanna}}, \bibnamefont{and}
  \bibinfo{author}{\bibfnamefont{M.}~\bibnamefont{Tissier}},
  \bibinfo{journal}{Phys. Rev. B} \textbf{\bibinfo{volume}{69}},
  \bibinfo{pages}{134413} (\bibinfo{year}{2004}).

\bibitem[{\citenamefont{Pawlowski}(2007)}]{pawlowski07}
\bibinfo{author}{\bibfnamefont{J.}~\bibnamefont{Pawlowski}},
  \bibinfo{journal}{Ann. Phys. (N.Y.)} \textbf{\bibinfo{volume}{322}},
  \bibinfo{pages}{2831} (\bibinfo{year}{2007}).

\bibitem[{\citenamefont{Rosten}(2012)}]{rosten12}
\bibinfo{author}{\bibfnamefont{O.}~\bibnamefont{Rosten}},
  \bibinfo{journal}{Phys. Rep.} \textbf{\bibinfo{volume}{511}},
  \bibinfo{pages}{177} (\bibinfo{year}{2012}).

\bibitem[{\citenamefont{Delamotte}(2012)}]{delamotte12}
\bibinfo{author}{\bibfnamefont{B.}~\bibnamefont{Delamotte}},
  \bibinfo{journal}{Lect. Notes Phys.} \textbf{\bibinfo{volume}{852}},
  \bibinfo{pages}{49} (\bibinfo{year}{2012}).

\bibitem[{\citenamefont{Canet et~al.}(2003)\citenamefont{Canet, Delamotte,
  Mouhanna, and Vidal}}]{canet03b}
\bibinfo{author}{\bibfnamefont{L.}~\bibnamefont{Canet}},
  \bibinfo{author}{\bibfnamefont{B.}~\bibnamefont{Delamotte}},
  \bibinfo{author}{\bibfnamefont{D.}~\bibnamefont{Mouhanna}}, \bibnamefont{and}
  \bibinfo{author}{\bibfnamefont{J.}~\bibnamefont{Vidal}},
  \bibinfo{journal}{Phys. Rev. B} \textbf{\bibinfo{volume}{68}},
  \bibinfo{pages}{064421} (\bibinfo{year}{2003}).

\bibitem[{\citenamefont{{G. De Polsi, I. Balog, M. Tissier and N.
  Wschebor}}(2020)}]{depolsi20}
\bibinfo{author}{\bibnamefont{{G. De Polsi, I. Balog, M. Tissier and N.
  Wschebor}}}, \bibinfo{journal}{Phys. Rev. E} \textbf{\bibinfo{volume}{101}},
  \bibinfo{pages}{042113} (\bibinfo{year}{2020}).

\bibitem[{\citenamefont{{I. Balog, H. Chaté, B. Delamotte, M. Marohni and N.
  Wschebor}}(2019)}]{balog19b}
\bibinfo{author}{\bibnamefont{{I. Balog, H. Chaté, B. Delamotte, M. Marohni and
  N. Wschebor}}}, \bibinfo{journal}{Phys. Rev. Lett.}
  \textbf{\bibinfo{volume}{123}}, \bibinfo{pages}{240604}
  (\bibinfo{year}{2019}).

\bibitem[{\citenamefont{Papenbrock and Wetterich}(1995)}]{papenbrock95}
\bibinfo{author}{\bibfnamefont{T.}~\bibnamefont{Papenbrock}} \bibnamefont{and}
  \bibinfo{author}{\bibfnamefont{C.}~\bibnamefont{Wetterich}},
  \bibinfo{journal}{Z. Phys. C} \textbf{\bibinfo{volume}{65}},
  \bibinfo{pages}{519} (\bibinfo{year}{1995}).

\bibitem[{\citenamefont{Morris and Tighe}(1999)}]{morris99}
\bibinfo{author}{\bibfnamefont{T.~R.} \bibnamefont{Morris}} \bibnamefont{and}
  \bibinfo{author}{\bibfnamefont{J.~F.} \bibnamefont{Tighe}},
  \bibinfo{journal}{J. High Energy Phys.} \textbf{\bibinfo{volume}{08}},
  \bibinfo{pages}{007} (\bibinfo{year}{1999}).

\end{thebibliography}
\end{document}